\documentclass[12pt]{article}
\pdfoutput=1
\usepackage{epsf,amsfonts,hyperref}
\usepackage[dvips]{color}
\usepackage{amsmath}
\usepackage{graphicx}
\renewcommand{\appendix}[1]{
    \setcounter{equation}{0}
    \renewcommand{\thesection}{\Alph{section}}
    \section{Appendix: \protect\indent #1}
}

\newcommand\encadremath[1]{\vbox{\hrule\hbox{\vrule\kern8pt
\vbox{\kern8pt \hbox{$\displaystyle #1$}\kern8pt}
\kern8pt\vrule}\hrule}}
\def\enca#1{\vbox{\hrule\hbox{
\vrule\kern8pt\vbox{\kern8pt \hbox{$\displaystyle #1$}
\kern8pt} \kern8pt\vrule}\hrule}}

\newcommand\figureframex[3]{
\begin{figure}[bth]
\hrule\hbox{\vrule\kern8pt
\vbox{\kern8pt \vbox{
\begin{center}
{\mbox{\epsfxsize=#1.truecm\epsfbox{#2}}}
\end{center}
\caption{#3}
}\kern8pt}
\kern8pt\vrule}\hrule
\end{figure}
}
\newcommand\figureframey[3]{
\begin{figure}[bth]
\hrule\hbox{\vrule\kern8pt
\vbox{\kern8pt \vbox{
\begin{center}
{\mbox{\epsfysize=#1.truecm\epsfbox{#2}}}
\end{center}
\caption{#3}
}\kern8pt}
\kern8pt\vrule}\hrule
\end{figure}
}

\makeatletter
\@addtoreset{equation}{section}
\makeatother
\newtheorem{theorem}{Theorem}[section]

\newtheorem{remark}{Remark}[section]
\newtheorem{proposition}{Proposition}[section]
\newtheorem{lemma}{Lemma}[section]
\newtheorem{corollary}{Corollary}[section]
\newtheorem{definition}{Definition}[section]
\def\br{\begin{remark}\rm\small}
\def\er{\end{remark}}
\def\bt{\begin{theorem}}
\def\et{\end{theorem}}
\def\bd{\begin{definition}}
\def\ed{\end{definition}}
\def\bp{\begin{proposition}}
\def\ep{\end{proposition}}
\def\bl{\begin{lemma}}
\def\el{\end{lemma}}
\def\bc{\begin{corollary}}
\def\ec{\end{corollary}}
\def\beaq{\begin{eqnarray}}
\def\eeaq{\end{eqnarray}}

\newcommand{\beq}{\begin{equation}}
\newcommand{\eeq}{\end{equation}}
\newcommand{\bea}{\begin{eqnarray}}
\newcommand{\eea}{\end{eqnarray}}

\newcommand{\td}[1]{{\tilde{#1}}}

\newcommand{\Pint}{{\int\kern -1.em -\kern-.25em}}

\newcommand\Res{\mathop{{\rm Res}}}

\begin{document}

\sloppy


\pagestyle{empty}
\baselineskip 16pt 
\begin{center}
\begin{Large}\fontfamily{cmss}
\fontsize{20pt}{30pt}
\selectfont
\medskip
\textbf{One-cut solution of the $\beta$-ensembles in the Zhukovsky variable}
\end{Large}\\
\bigskip
\bigskip
\begin{large} 
 {O. Marchal}$^{\dagger}$\footnote{olivier.marchal@polytechnique.org}
\end{large}
\\
\bigskip
\begin{small}
$^{\dagger}$ {\em Department of Mathematics and Statistics\\
University of Alberta\\
Edmonton, Canada\\
}
\end{small}
\end{center}

\author{O. Marchal}

\bigskip
\smallskip
\bigskip

\begin{center}{\bf Abstract}
\end{center}
\smallskip

In this article, we study in detail the modified topological recursion of the one matrix model for arbitrary $\beta$ in the one cut case. We show that for polynomial potentials, the recursion can be computed as a sum of residues. However the main difference with the hermitian matrix model is that the residues cannot be set at the branchpoints of the spectral curve but require the knowledge of the whole curve. In order to establish non-ambiguous formulas, we place ourselves in the context of the globalizing parametrization which is specific to the one cut case (also known as Zhukovsky parametrization). This situation is particularly interesting for applications since in most cases the potentials of the matrix models only have one cut in string theory. Finally, the article exhibits some numeric simulations of histograms of limiting density of eigenvalues for different values of the parameter $\beta$.

\tableofcontents

\vspace{26pt}
\pagestyle{plain}
\setcounter{page}{1}


\section{Introduction}

Recently, the question of arbitrary $\beta$ matrix models, also known as $\beta$-ensembles, has been put forward for its possible applications to a $\beta$-deformed Chern-Simons theory \cite{Marino}. Moreover, if a lot of results are known for the case of hermitian matrix models ($\beta=1$), like for example: the famous topological recursion introduced by Eynard and Orantin \cite{OE} or the connection with orthogonal polynomials \cite{Mehta}, much less is known for real-symmetric matrices ($\beta=\frac{1}{2}$) or self-dual quaternionic matrices ($\beta=2$), even if a reformulation in terms of skew-orthogonal polynomials is known (\cite{Mehta}). Moreover, when $\beta$ is arbitrary, no connection with orthogonal polynomials is explicitly known and the knowledge about these models remain very partial. Nevertheless, these $\beta$-ensembles can be studied by the computation of their corresponding correlation functions and one can write some deformed loop equations in the same way as for the standard hermitian matrices. However, it is also known that solving these loop equations is much more difficult when $\beta\neq 1$ since in this case they do not remain algebraic but includes a derivative term. In various articles (\cite{MoiBertrand, MoiLeonidBertrand, MoiLeonidBertrand2}), it was shown that under a specific regime, namely when $\beta$ is connected to the number of eigenvalues $N$ by $N\beta=\text{Cste}$ then one can recover a ``quantum" version of the topological recursion for the hermitian case, with the main difference that the curve is no longer algebraic but a differential equation. However, if one wants to study a specific and fixed value of the exponent $\beta$, then one has to solve the refined loop equations where $\beta$ remains fixed. In this context, the only known way is to adapt the standard topological recursion to a ``refined" topological recursion. This method is not new, since it was the first one used to deal with the $\frac{1}{N}$ expansion even for the hermitian case before Eynard and Orantin found a important reformulation of this method in the topological recursion (which is indeed more convenient since it only uses local properties of the spectral curve) and the introduction of symplectic invariants for any algebraic curve \cite{OE}. In this article, we first remind the reader how to determine the loop equations in the context of arbitrary $\beta$ one-matrix models. Then we place ourselves in the case of a one-cut solution and we use the existing globalizing parametrization to solve the recursion. All these results are presented in great details, making the present paper self-contained and possibly adaptable for more general considerations like the multi-cut case. We note also that our method is closely related to the method used by Brini, Mari\~{n}o and Stevan in \cite{Marino}, with the main difference that our parametrization leads to non-ambiguous formulas (no square roots) and to a much easier implementation on a computer. Some similar results and simulations can be found in \cite{Nadal} for the case of the Gaussian potential (for which the partition function is already known). Finally, to illustrate our results and to exhibit the importance of the parameter $\beta$, we present some simulations of limiting density of eigenvalues for the case of an even-quartic potential with different values of $\beta$.

\section{Notation and determination of the spectral curve}

In this article, we are interested in computing the first terms for the Chekhov-Eynard recursion for matrix models for arbitrary value of $\beta$. Before deriving the loop equations, we need to fix the notations used in this article. In our case we are interested in matrix models of the form:
\beq Z_{N,\beta}=\frac{1}{N! (2\pi)^N} \int_{\mathbb{R}^N} d\lambda_1 \dots d\lambda_N |\Delta(\lambda)|^{2\beta} e^{-\frac{N \beta}{t_0} \underset{i=1}{\overset{N}{\sum}} V(\lambda_i)}\eeq
where $V(x)$ is a potential, which in this article is supposed to be a polynomial of degree $d=4$: $V(x)=\frac{1}{4}x^4+\frac{t_3}{3}x^3+ \frac{t_2}{2}x^2+t_1x$ with only one local minimum on the real axis (one-cut assumption). The notation $\Delta(\lambda)$ stands for the Vandermonde determinant: $\Delta(\lambda)=\underset{i<j}{\overset{N}{\prod}} (\lambda_i-\lambda_j)$ For people interested in topological string theory, we remind the reader that the connection with the string coupling constant is $g_s=\frac{t_0}{N}$ so that a series expansion in $\frac{1}{N}$ can also be interpreted as a series expansion in $g_s$. Moreover, as usual in matrix models, we introduce the connected correlation functions:

\bea \label{TTT} W(x)&=&\frac{t_0}{N}\left< \sum_{i=1}^N \frac{1}{x-\lambda_i}\right>\cr
W(x,y)&=&\left(\frac{t_0}{N}\right)^2\left< \sum_{i,j=1}^N \frac{1}{x-\lambda_i}\frac{1}{y-\lambda_j}\right>_c\cr
W(x_1,\dots,x_n)&=&\left(\frac{t_0}{N}\right)^n \left< \sum_{i_1,\dots,i_n}^N  \frac{1}{x_1-\lambda_{i_1}}\dots \frac{1}{x_n-\lambda_{i_n}}\right>_c
\eea

The bracket notation $<f(\lambda_1,\dots,\lambda_N)>$ stands for the expected value of the function $f$:
\beq \left< f(\lambda_1,\dots,\lambda_N) \right>=\frac{1}{Z_{N,\beta}} \frac{1}{N! (2\pi)^N} \int_{\mathbb{R}^N} d\lambda_1 \dots d\lambda_N |\Delta(\lambda)|^{2\beta} f(\lambda_1,\dots,\lambda_N) e^{-\frac{N \beta}{t_0} \underset{i=1}{\overset{N}{\sum}} V(\lambda_i)}\eeq
The index $_c$ stands for the connected component of the correlation function:
\bea <A_1> &=& <A_1>_c \cr
<A_1 A_2> &=& <A_1 A_2>_c + <A_1>_c <A_2>_c \cr
<A_1 A_2 A_3> &=& <A_1 A_2 A_3>_c + <A_1 A_2>_c <A_3> + <A_1 A_3>_c <A_2> \cr
&&+ <A_2 A_3>_c <A_3> + <A_1> <A_2> <A_3>\cr
<A_1\dots A_n>&=&<A_J>=\sum_{k=1}^n\sum_{I_1 \bigsqcup I_2 \dots \bigsqcup I_k=J}\prod_{i=1}^k <A_{I_i}>_c\cr 
\eea

Finally, in order to close the future loop equations, we also need to introduce the following functions:
\bea P(x)&=&\frac{t_0}{N}\left< \sum_{i=1}^N\frac{V'(x)-V'(\lambda_i)}{x-\lambda_i}\right>\cr
 P(x_1,\dots,x_n)&=&\left(\frac{t_0}{N}\right)^n \left< \sum_{i_1,\dots i_n=1}^N \frac{V'(x)-V'(\lambda_{i_1})}{x-\lambda_{i_1}}\frac{1}{x-\lambda_{i_2}} \dots \frac{1}{x-\lambda_{i_n}} \right>\cr
 \eea
These functions share the important property to be polynomials of degree $2$ in their first variable.   

\section{Loop equations for arbitrary $\beta$}

\subsection{First loop equation}

The loop equations are a infinite set of equations connecting the correlation functions and the functions $P(x_1,\dots x_n)$. They come from the invariance by reparametrization of the integral. These equations are well known for the hermitian case ($\beta=1$) and can be found in many different places in the literature. For arbitrary $\beta$, the equations are somewhat less known and can be obtained in many different ways. We choose here to remind briefly a possible and completely rigorous way to derive them. In particular this method has the advantage that one can easily include hard edges, that is to say restrict the integration domain not to $\mathbb{R}$ but to a fixed interval (the extremities being called hard edges). The trick is to look at the following total derivative:

\beq I_p=\frac{1}{Z_{N,\beta}}\int d\lambda_1\dots d\lambda_N \frac{\partial}{\partial \lambda_p}\left(\frac{1}{x-\lambda_p} e^{-\frac{N\beta}{t_0}\underset{i=1}{\overset{N}{\sum}} V(\lambda_i)} |\Delta(\lambda)|^{2\beta}\right)=0\eeq

The result here is zero since it is a total derivative and the contour of integration has no hard edge (i.e. goes from $-\infty$ to $\infty$). This integral gives three contributions depending on which term of the product the derivative operator acts. The first one is given by 
$$\left< \frac{1}{(x-\lambda_p)^2}\right>$$
The second one is:
$$-\frac{N\beta}{t_0}\left< \frac{V'(\lambda_p)}{x-\lambda_p}\right>=\frac{N\beta}{t_0}\left< \frac{V'(x)-V'(\lambda_p)}{x-\lambda_p}\right> -\frac{N\beta}{t_0}V'(x) \left< \frac{1}{x-\lambda_p}\right> $$
Finally, the last one involves the logarithmic derivative of the Vandermonde determinant:
\beq 2\beta \left<\frac{1}{x-\lambda_p}\frac{\partial_{\lambda_p}|\Delta(\lambda)|}{|\Delta(\lambda)|}\right>=2\beta \left<\frac{1}{x-\lambda_p}\sum_{j \neq p} \frac{1}{\lambda_j-\lambda_p}\right>\eeq
When summing over $p$ one can antisymmetry of the last term:
\bea &&\sum_{p=1}^N\sum_{j\neq p}\frac{1}{x-\lambda_p}\frac{1}{\lambda_j-\lambda_p}= \frac{1}{2} \sum_{p, j\neq p}^N \left(\frac{1}{x-\lambda_p}\frac{1}{\lambda_j-\lambda_p}- \frac{1}{x-\lambda_j}\frac{1}{\lambda_j-\lambda_p}\right)\cr
&&=\frac{1}{2} \sum_{p, j\neq p}^N \frac{1}{(x-\lambda_p)(x-\lambda_j)}\cr
\eea
Inserting all these results together leads to the first loop equation: 
\bea \label{FirstLoop} &&0=\frac{1}{Z_{N,\beta}}\sum_{p=1}^N\int d\lambda_1\dots d\lambda_N \frac{\partial}{\partial \lambda_p}\left(\frac{1}{x-\lambda_p} e^{-\frac{N\beta}{t_0}\underset{i=1}{\overset{N}{\sum}} V(\lambda_i)} |\Delta(\lambda)|^{2\beta}\right)\cr
&&=\frac{N(\beta-1)}{t_0}W'(x) +\frac{N\beta}{t_0} W^2(x)-\frac{N^2\beta}{t_0^2} V'(x)W(x)+\frac{N^2\beta}{t_0^2} W(x,x)+\frac{N^2\beta}{t_0^2}P(x)\cr \eea

As mentioned before, this method to get the loop equations is useful if one has to face hard edges. Indeed in this case, the total integral can be computed and is non-zero. For example, if the integral only goes from $a$ to $+\infty$ then we find:
\beq -\sum_{p=1}^N \frac{e^{-\frac{N\beta}{t_0}V(a)}}{x-a}\frac{1}{Z_{N,\beta}}\int_{[a,+\infty]^{N-1}} d\lambda_i e^{-\frac{N\beta}{t_0}\underset{i\neq p}{\overset{N}{\sum}} V(\lambda_i)}|\Delta(\lambda, \lambda_p=a)|^{2\beta}\eeq
Which gives:
\beq 
(1-\frac{1}{\beta})\frac{t_0}{N}\partial_z W(z) + W(z)^2 -V'(z)W(z)+W(z,z)= -P(z)+ \frac{c(N,\beta)}{z-a}\eeq
with an unknown function $c(N,\beta)$. This gives us the fact that the one-point function has a simple pole at the hard edge, a fact that is well known in this situation \cite{Hardedges, Hardedges2}.

\subsection{Other loop equations}

To get the other loop equations, one can use the following total derivative:
\bea &&0=I_{p,i_2,\dots,i_n}(x_2,\dots,x_n)=\cr
&&\frac{1}{Z_{N,\beta}}\int d\lambda_1\dots d\lambda_N \frac{\partial}{\partial \lambda_p}\Big(\left(\frac{1}{x-\lambda_p}\frac{1}{x_2-\lambda_{i_2}}\dots \frac{1}{x_n-\lambda_{i_n}}\right)e^{-\frac{N\beta}{t_0}\underset{i=1}{\overset{N}{\sum}} V(\lambda_i)} |\Delta(\lambda)|^{2\beta}\Big)\cr\eea
and use the same kind of decomposition as before. The result has been known for a long time (and the derivation can be found in \cite{Mehta, Nadal}) and give us the fact that the correlation functions satisfy the following loop equations:
\bea \label{SuperLoop} &&\left(2 W(x_1)-V'(x_1)+\hbar\gamma\right)W(x_1,\dots,x_p)+ \hbar^2 W(x_1,x_1,x_2,\dots,x_p)\cr
&& +\sum_{j=1}^{p-2}\sum_{I \in K_j} W(x_1,x_I)W(x_1,x_{K/I})\cr
&&=P(x_1,\dots,x_p)-\sum_{j=2}^k \frac{\partial}{\partial x_j} \frac{ W(x_2,\dots,x_j,\dots,x_p)- W(x_2,\dots,x_1,\dots,x_p)}{x_j-x_1}\cr
\eea 
which are valid $\forall k \geq 2$ and where the notation $K_j$ represents all subsets of $\{ x_1,x_2,\dots, x_p \}$ with $j$ distinct variables. Note that the r.h.s. is polynomial in $x_1$ and has no singularity when $x_1\to x_j$.
For $k=1$, equation \eqref{FirstLoop} can be rewritten into a more compact form:
\beq \encadremath{ W^2(x)-V'(x)W(x)+\hbar^2 W(x,x)+\hbar \gamma W(x)=-P(x)}\eeq
where the coefficients $\hbar$ and $\gamma$ are given by:
\beq \encadremath{ \hbar= \frac{t_0}{N \sqrt{\beta}} \,\,\,\, ,\,\,\,\, \gamma=\sqrt{\beta}-\sqrt{\beta^{-1}} }\eeq

Note that $\hbar$ is typically small for $N$ large and that a series expansion in $\frac{1}{N}$ is completely equivalent to a series expansion in $\hbar$ up to trivial factors. When $\beta$ is fixed, which is the case studied in this article, the parameter $\gamma$ is also fixed and so is just used here for convenience to simplify the formulas.
As mentioned in the introduction, on can also study the case where $$\sqrt{\beta} -\sqrt{\beta^{-1}} \sim N$$ i.e. a case where $\beta$ is not fixed. Then $\gamma$ would be of order $N$ leading to completely different solutions to the loop equations in relation with a Schr\"{o}dinger equation as presented in the articles \cite{MoiBertrand,MoiLeonidBertrand, MoiLeonidBertrand2} having some connections with the AGT conjecture \cite{AGT,Recent}.
Finally, note that the case of hermitian matrix models corresponds to $\gamma=0$ in which the derivative term in the loop equations disappears giving a set of purely algebraic equations. 

\section{Large $N$ expansion of the correlation functions and one-cut assumption}

Since the potential $V(x)$ is assumed to have only one local extremum on the real line (where the integral is taken), it is natural to think that the eigenvalues will accumulate around this minimum and in the limit will condense to a density distribution given by a unique interval around this minimum. This gives rise to a limiting density of eigenvalues $\rho_\infty(x)$ which is supported on an interval $[a,b]$ containing the minimum of the potential. The fact that the support of the distribution only has a unique interval is often called the one-cut case in matrix models. Note that this is a very specific case since in general, a potential may have various minima along the path of integration. Such situations would lead to multi-cut solutions of the loop equations, in which the fraction (called ``filling fraction") of eigenvalues on each interval has to be fixed either dynamically (equality of chemical potentials also known as convergent matrix models or statically (fixed filling fraction) also known as formal matrix models).
In this article, we assume that the potential belongs to the one-cut case. This situation has the advantage of being easier (it does not require abstract algebraic geometry tools) and in the hermitian case it has been proved that under this assumption, the correlation functions admit a series expansion in $\frac{1}{N}$ in \cite{Referee1}. For multi-cut case models, the difference between convergent matrix model and formal matrix model is subtle and has been studied only in details in the case of hermitian matrix models \cite{MultiCutCase, MultiCutCase2}. In particular, it is no longer true that the correlation functions have a $\frac{1}{N}$ expansion and one has to integrate over the filling fractions. For arbitrary $\beta$, there is no rigorous proof that the correlation functions admit a series expansion in $\frac{1}{N}$ but since there is only one saddle point in the model, it makes it reasonable to assume that the situation cannot be different. Therefore we assume in the rest of the article that the correlation functions admit the following large $N$ expansion (rewritten with the help of the $\hbar=\frac{t_0}{N \sqrt{\beta}}$ variable):

\bea W(x_1,\dots,x_p)&=&\sum_{g=0,1/2}^\infty \hbar^{2g} W_g(x_1,\dots,x_p)\cr
P(x_1,\dots,x_p)&=&\sum_{g=0,1/2}^\infty \hbar^{2g} P_g(x_1,\dots,x_p)\eea 

Note that the sums are taken with $g$ to be an integer or the half of an integer. In the case of hermitian matrix model, since the loop equations only involves $\hbar^2$ (and not directly $\hbar$), then only integer values of $g$ are non-zero, but for arbitrary $\beta$, i.e. when $\gamma$ is not zero, then the loop equations have a term involving $\hbar$ and thus the series expansion also contains odd powers of $\hbar$.

Moreover, since the loop equations involve the parameter $\gamma$, it is useful to write down a series expansion also in powers of $\gamma$. From the shape of the loop equations, it is easy to see that for a given $W_g$ or $P_g$ the expansion is only finite. Indeed the first loop equation (after projection at zero order in $\hbar$) is independent from $\gamma$ (which appears with a $\hbar$ term in front) so $W_0(x)$ will be independent from $\gamma$. Then a simple recursion gives us that only a finite numbers of powers of $\gamma$ can be involved for each correlation function. More precisely we have:

\bea  
W(x_1,\dots,x_p)&=&\sum_{g=0,1/2}^\infty \hbar^{2g} W_g(x_1,\dots,x_p)=\sum_{g=0,1/2}^\infty\sum_{k=0}^{[g]} \hbar^{2g}\gamma^{2g-2k} W_{k,2g-2k},(x_1,\dots,x_p)\cr
P(x_1,\dots,x_p)&=&\sum_{g=0,1/2}^\infty \hbar^{2g} P_g(x_1,\dots,x_p)=\sum_{g=0,1/2}^\infty\sum_{k=0}^{[g]} \hbar^{2g}\gamma^{2g-2k} P_{k,2g-2k},(x_1,\dots,x_p)\cr
\eea

It is also interesting to do the same kind of expansion for the free energy:
\beq \log(Z_{N,\beta})=\mathcal{F}=\sum_{g=0}^\infty \hbar^{2g}F_g=\sum_{g=0,1/2}\sum_{k=0}^{[g]} \hbar^{2g} \gamma^{2g-2k}F_{k,2g-2k}\eeq
which gives:
\bea F_{0}&=&F_{0,0}\cr
F_{1/2}&=&\gamma F_{0,1}\cr
F_{1}&=&F_{1,0}+\gamma^2 F_{0,2}\cr
F_{3/2}&=&\gamma F_{1,1}+\gamma^3 F_{0,3}\cr
F_{2}&=& F_{2,0}+\gamma^2F_{1,2}+\gamma^4 F_{0,4}\cr
\eea
and more generally:
\bea F_{g}&=&\sum_{i=0}^gF_{g-i,2i}\gamma^{2i}  \cr
F_{g/2}&=&\gamma \sum_{i=0}^{[g/2]}F_{g/2-1/2-i,2i+1}\gamma^{2i}\eea

Usually, since $\gamma$ is finite, one is more interested in the numbers $F_g$ and the functions $W_g(x_1,\dots,x_p)$ that have a physical meaning since they represent the corrections at large $N$. Fortunately, as we can see, we can extract them just by finite sum of the $F_{k,l}$ and of the $W_{k,g}(x_1,\dots,x_p)$. Here, the introduction of the second expansion in $\gamma$ is used for computational convenience and to make some correspondence with various articles in the literature \cite{Nadal, Marino, ChekEynbeta, Correct}, but we stress that the main goal is to obtain the $F_g$'s and the $W_g(x_1,\dots,x_p)$'s.

\section{Solving the loop equations}

\subsection{Finding the resolvent}

The work from Chekov and Eynard \cite{ChekEynbeta} gives us the way to solve the previous loop equations. Indeed, after projecting the first loop equation onto the order $\hbar^0$ one finds the algebraic equation:
\beq W_0^2(x)-V'(x)W_0(x)=-P_0(x)\eeq
where $W_0(x)$ is often called in the literature as ``the resolvent". Since $P_0(x)$ is a polynomial, one find the solution:
\beq W_0(x)= \frac{V'(x)}{2} - \sqrt{\frac{V'(x)^2}{4}-P_0(x)}\eeq
The minus sign is chosen so that $W_0(x)$ behaves like $O\left(\frac{1}{x}\right)$ at infinity as expected from the definition of $W(x)$ given by \eqref{TTT}.

It is also usual to introduce the function $y(x)$ often called the ``spectral curve" of the model:
\beq y(x)=W_0(x)-\frac{V'(x)}{2} \,\,\, ,\,\,\, y^2(x)=U(x)=\frac{V'(x)^2}{4}-P_0(x)\eeq
Assuming the one cut solution corresponds to assume that the solution is of the form:
\beq y(x)= M(x)\sqrt{(x-a)(x-b)}= \left(m_2x^2+m_1x+m_0\right)\sqrt{(x-a)(x-b)}\eeq
where $M(x)$ is a polynomial of degree $2=d-2$. Moreover, we also assume that the case at stake is regular (and not critical), meaning the the polynomial $M(x)$ has no zeros in the interval $[a,b]$. The coefficients of $M(x)$ and the endpoints $a$ and $b$ are fully determined by the behavior at infinity:
\beq y(x)= \frac{V'(x)}{2}+ \frac{t_0}{x}+ O\left(\frac{1}{x^2}\right)\eeq
which comes directly from the definition of $W(x)$. In practice, for an arbitrary potential, it is easy to write down the relation between the coefficients of $M(x)$ and the endpoints $a$ and $b$. But in general, the equations giving the endpoints $a$ and $b$ as a function of the coefficients of the potentials $t_i$ and $t_0$ become rapidly highly complicated. In our case, if we write the potential as:
\beq V(x)=\frac{t_4}{4}x^4+ \frac{t_3}{3}x^3+ \frac{t_2}{2}x^2+t_1x\eeq
We have the simple relations:
\bea \label{m_i} m_2&=&-\frac{t_4}{2}\cr
m_1&=&-\frac{(a+b)t_4}{4}-\frac{t_3}{2}\cr
m_0&=&-\frac{t_4}{16}\left(3a^2+3b^2+2ab\right)-\frac{t_2}{2}- t_3\frac{a+b}{4}\cr
\eea 
But the connection with the endpoints is highly non-trivial since they satisfy the following set of algebraic equations:
\bea\label{ab} 0&=&2t_3\left(2ab+3a^2+3b^2\right)+8t_2\left(a+b\right)+16t_1 +t_4\left(3a^2b +3ab^2+5a^3+5b^3\right)\cr
-256t_0&=&16t_3\left(a^2b+ab^2-a^3-b^3\right) -16t_2\left(a^2+b^2\right) +t_4\left(6a^2b^2 +12a^3b+12ab^3-15\left(a^4+b^4\right)\right)\cr
\eea
This set of algebraic equations cannot be solved analytically but can easily be handled numerically. Note that both equations are symetric functions of $a$ and $b$, and that in order to have a proper interpretation, we will assume $a<b$.

\subsection{Computing $W_{1/2}(x)$}

As soon as the resolvent has been determined, one can solve the loop equations and find the higher correlation functions by recursion. In the case of arbitrary $\beta$, the formalism was introduced by Chekhov and Eynard in \cite{ChekEynbeta} and slightly corrected in \cite{Correct}. We summarize here the results. First we can rewrite the loop equations projected on every power of $\hbar$. This gives the following projected loop equations (we note $x_K=\{x_2,\dots,x_n\}$ for compactness): 
\bea \label{Projected loop} 2y(x_1)W_g(x_1,x_K)&=& -\sum_{h=0}^g\sum_{I \subset K}^{'} \mathcal{W}_h(x_1,x_I)\mathcal{W}_{g-h}(x_1,x_{K/I}) -W_{g-1}(x_1,x_1,x_K)\cr
&&-\gamma \partial_{x_1} W_{g-\frac{1}{2}}(x_1,x_K)+U_g(x_1,x_K)
\eea
where the sum $\underset{h=0}{\overset{g}{\sum}}\underset{I \subset K}{\overset{'}{\sum}}$ means that we exclude the two case $(h=0, I=\emptyset)$ and $(h=g,I=K)$ which are already included in the l.h.s. Also note that $x_1\mapsto U_g(x_1,x_K)$ is regular on the cut since it is a combination of $P_g(x_1,x_K)$ and $\underset{j=2}{\overset{k}{\sum}} \frac{\partial}{\partial x_j} \frac{ W_g(x_2,\dots,x_j,\dots,x_p)}{x_j-x_1}$ which are both regular functions of $x_1$ on the cut.

We can also project these loop equations on the second series expansion in $\gamma$. This leads us to:
\bea 2y(x_1)W_{g,l}(x_1,x_K)&=& -\sum_{g_1,l_1\geq0, l_1+g_1>0} \sum_{h=0}^g\sum_{I \subset K}^{'}\mathcal{W}_{g_1,l_1}(x_1,x_I)\mathcal{W}_{g-g_1,l-l_1}(x_1,x_{K/I})\cr
&&-W_{g-1,l}(x_1,x_1,x_K)-\partial_{x_1}W_{g-\frac{1}{2}, l-1}(x_1,x_K)+U_{g,l}(x_1,x_K)\cr
\eea
where again $x_1\to U_{g,l}(x_1,x_K)$ is regular on the cut.

In this recursion, we have introduced the shifted two-points function:
\beq \encadremath{ \forall\,  g\geq1: \,\, \, \mathcal{W}_g(x_1,\dots,x_p)=W_g(x_1,\dots,x_p)+\frac{1}{2}\frac{\delta_{g=0}\delta_{p=2}}{(x_1-x_2)^2} } \eeq
Observe that $W_0(x_1,x_2)$ is regular at $x_1=x_2$ since it is a correlation function. But the shifted function has a double pole without residue when $x_1 \to x_2$. The necessity of the shift is to take care properly of the terms 
$$\sum_{j=2}^n\frac{\partial}{\partial x_j} \frac{ W_g(x_2,\dots,x_j,\dots,x_n)-W_g(x_1,\dots,x_1,\dots,x_n)}{x_j-x_1}$$
present in the loop equations \eqref{SuperLoop}.

Then one observes that for arbitrary $\beta$ the functions $W_{g}(x_1,\dots,x_p)$ have two kind of singularities. First they have square-root singularities coming by recursion from the one of the spectral curve. But when $\beta\neq 1$, the initialization of the recursion with $W_{1/2}(x)$ gives us also poles at the other zeros of $y(x)$, that is to say zeros of the moment function $M(x)$. For $\beta=1$ this does not happen since the recursion is initialized with $W_0(x_1,x_2)$ which is known in the one cut case to be just $\frac{1}{2(x_1-x_2)^2}$ or more generally connected to the fundamental bi-differential (In the literature this kernel is sometimes erroneously referred to as the "Bergman" kernel) for the multi-cut case. However, for arbitrary $\beta$ the solutions can still be rewritten as contour integral around the cut by the following formulas \cite{Marino}:

\beq W_{1/2}(x_1)=-\frac{\gamma}{4\pi i\sqrt{(x_1-a)(x_1-b)}} \oint_{\mathcal{C}_{[a,b]}}\frac{dx}{M(x)(x_1-x)}W'_0(x)\eeq
\beq \label{W_{1/2}} W_{1/2}(x_1,x_2)=-\frac{1}{4i\pi\sqrt{(x_1-a)(x_1-b)}} \oint_{\mathcal{C}_{[a,b]}} \frac{dx}{M(x)(x_1-x)} \Big( 2 W_{1/2}(x)\mathcal{W}_0(x,x_2)+\gamma \partial_x W_0(x,x_2) \Big)\eeq
\beq \label{W_{1}} W_{1}(x_1)=-\frac{1}{4i\pi\sqrt{(x_1-a)(x_1-b)}} \oint_{\mathcal{C}_{[a,b]}} \frac{dx}{M(x)(x_1-x)} \Big( W_0(x,x)+\gamma \partial_x W_{1/2}(x) \Big)\eeq
\bea \label{recursion} W_g(x_1,x_K)&=& -\frac{1}{4i\pi\sqrt{(x_1-a)(x_1-b)}} \oint_{\mathcal{C}_{[a,b]}}\frac{dx}{M(x)(x_1-x)} \Big( \sum_{h=0}^g \sum_{J \subset K}^{'} \mathcal{W}_h(x,x_J) \mathcal{W}_{g-h}(x,x_{K/J}) \cr
&&+ W_{g-1}(x,x,x_K) +\gamma \partial_x W_{g-1/2}(x,x_K) \Big)\cr
\eea
where again $K$ is a generic notation for the set $x_K=\{ x_1,\dots, x_k\}$ and the sum does not include $(h=0,J=\emptyset)$ nor $(h=g, J=K)$. The contour of integration $\mathcal{C}_{[a,b]}$ is a contour circling the interval $[a,b]$ in the counterclockwise direction. In the case when $M(x)$ is a polynomial, one can deform the contour and circle the other singularities of the integrand to get explicit formulas. 

For example, we find that:
\bea W_{1/2}(x_1)&=& \frac{3\gamma}{2\sqrt{(x_1-a)(x_1-b)}} -\frac{\gamma}{4} \frac{2x_1-a-b}{(x_1-a)(x_1-b)}\cr && -\frac{\gamma}{2\sqrt{(x_1-a)(x_1-b)}}\sum_{i\in \{\pm\} } \frac{\sqrt{(x_1-a)(x_1-b)} -\sqrt{(x_i-a)(x_i-b)}}{x_1-z_i}\cr\eea
in accordance with formula $(2.54)$ of \cite{Marino}. Here, we have noted $x_{\pm}$ the two zeros of $M(x)$ but it is trivial to generalize such formula for an arbitrary but finite number of zeros of $M(x)$.

However, as we will see in the next sections, using the $x$ variable is rather awkward since one always has to specify which branch of the square root is being used. In particular, to get a correct answer, one has to take the branches such that $\sqrt{\sigma(x_-)}=-\sqrt{\sigma(x_+)}$ since as we will see later, both $x_\pm$ lie in different sheets. This can be understood easily for $W_{1/2}(x_1)$ but will become much harder for the next correlation functions. Finally, as noted in \cite{Marino}, the last formula shows that the correlation function $W_{1/2}(x)$ has no singularity at the zeros of $M(x)$ and behaves like $O\left(\frac{1}{x^2}\right)$ at infinity, which was expected since it is a correlation function.

In our example of a quartic potential, we can compute explicitly the zeros of $M(x)$:
\beq 
x_{\pm}=-\frac{t_3}{2t_4}-\frac{a+b}{4}\pm\frac{1}{4}\sqrt{4\left(\frac{t_3}{t_4}\right)^2-4t_3(a+b)-2ab-16\frac{t_2}{t_4}-5(a^2+b^2)}\eeq

\subsection{Computing $W_{0}(x_1,x_2)$}

From the loop equations, it is easy to observe that the leading order of the two-point functions $W_{0}(x_1,x_2)$ is independent of $\beta$. Therefore, it is given by the same formula as for the hermitian matrix model and can be found in different places in the literature (for example equation $6.9$ in \cite{Correct}):

\bea \label{ouh} W_0(x_1,x_2)&=&\frac{1}{4\sqrt{\sigma(x_1)}\sqrt{\sigma(x_2)}}\left( \left(\frac{\sqrt{\sigma(x_1)}-\sqrt{\sigma(x_2)}}{x_1-x_2}\right)^2-1\right)\cr
&=&-\frac{1}{2(x_1-x_2)^2} +\frac{2x_1x_2 -(a+b)(x_1+x_2)+2ab}{4 (x_1-x_2)^2\sqrt{(x_1-a)(x_1-b)}\sqrt{(x_2-a)(x_2-b)}}\cr\eea
Note that it is indeed regular at $x_1=x_2$ as expected. It is not the case of the shifted version:
\beq  \encadremath{ \mathcal{W}_0(x_1,x_2)=\frac{2x_1x_2 -(a+b)(x_1+x_2)+2ab}{4 (x_1-x_2)^2\sqrt{(x_1-a)(x_1-b)}\sqrt{(x_2-a)(x_2-b)}} }\eeq

\subsection{Recursion in the $x$ variable}

One can iterate the contour deformation for the next correlation functions. Most of the results presented here can be found in \cite{Marino}, but for completeness, we present them shortly in this section. First one can compute $W_1(x_1)$: the integrand $\frac{dx}{M(x)(x_1-x)} \Big( W_0(x,x)+\gamma \partial_x W_{1/2}(x) \Big)$ only has pole at $x=x_1$ and $x=x_{\pm}$ since the behavior at infinity of the correlation functions give us that there is no pole there. At $x=x_1$ only $\frac{1}{x-x_1}$ is singular and at $x=z_{\pm}$ only $\frac{1}{M(x)}$ is singular. So the residues are straightforward to compute and we find:
\beq \encadremath{ W_1(x_1)= -\frac{1}{2y(x_1)}\left( W_2(x_1,x_1)+\gamma W'_{1/2}(x_1)\right)+ \frac{1}{2\sqrt{\sigma(x_1)}}\sum_{i\in \{\pm\}} \frac{W_0(x_i,x_i)+\gamma W'_{1/2}(x_i)}{M'(x_i)(x_1-x_i)}  }
\eeq

The case of $W_{1/2}(x_1,x_2)$ is a little bit more complicated. Indeed, the integrand is:
$$\frac{dx}{M(x)(x_1-x)} \Big( 2 W_{1/2}(x)\mathcal{W}_0(x,x_2)+\gamma \partial_x W_0(x,x_2) \Big)$$
There is again a simple pole at $x=x_1$ and simple poles at $x=x_{\pm}$. But this time, we have also an additional pole at $x=x_2$ coming from $\mathcal{W}_0(x,x_2)\sim \frac{1}{2(x-x_2)^2}$. Fortunately, this pole is not present in the derivative term $\gamma \partial W_0(x,x_2)$ since it involves only the non-shifted two-points function. Therefore, we find:

\bea
W_{1/2}(x_1,x_2)&=&-\frac{1}{2y(x_1)}\Big( 2 W_{1/2}(x_1)\mathcal{W}_0(x_1,x_2)+\gamma \partial_{x_1} W_0(x_1,x_2) \Big) \cr
&&+ \frac{1}{2\sqrt{\sigma(x_1)}}\sum_{i \in \{\pm\}} \frac{1}{(x_1-x_i)M'(x_i)}\left(2W_{1/2}(x_i)\mathcal{W}_{0}(x_i,x_2)+\gamma \left(\partial_x W_0(x,x_2)\right)_{|x=x_i} \right)\cr
&&+\frac{1}{4\sqrt{\sigma(x_1)}}\Big( \frac{W'_{1/2}(x_2)}{M(x_2)(x_1-x_2)} -\frac{W_{1/2}(x_2)M'(x_2)}{M(x_2)^2(x_1-x_2)} +\frac{W_{1/2}(x_2)}{M(x_2)(x_1-x_2)^2}\Big)
\eea

The recursion \eqref{recursion} goes exactly in the same way. The integrand only have simple poles at $x=x_{\pm}$ coming from $M(x)$ in the denominator, since the functions $W_g(x,x_J)$ will be regular at these two points. At infinity, the residue is always null because all functions $\mathcal{W}_h(x,x_J)$ involved in the recursion behave like $O\left(\frac{1}{x^2}\right)$ at infinity. The recursion scheme is also stable with the kind of singularities, meaning that the integrand always only have a cut on $[a,b]$, poles at $x=x_{\pm}$, $x=x_2$ and at $x=x_i$. These residues at $x=x_{\pm}$ can be evaluated easily since the simple poles only come from $\frac{1}{M(x)}$. The pole at $x=x_1$ is simple coming from $\frac{1}{x_1-x}$. The poles at $x=x_i$ come from $\mathcal{W}_0(x,x_i)\sim \frac{1}{2(x-x_i)^2}$:

\bea W_g(x_1,x_K)&=& \frac{1}{4\sqrt{\sigma(x_1)}}\sum_{i=2}^k\frac{1}{M(x_i)(x_1-x_i)}\Big( \frac{W'_{g}(x_2,\dots,x_k)}{M(x_i)(x_1-x_i)} -\frac{W_{g}(x_2,\dots,x_k)M'(x_i)}{M(x_i)^2(x_1-x_i)}\cr &&+\frac{W_{g}(x_2,\dots,x_k)}{M(x_i)(x_1-x_i)^2}\Big)\cr
&&+\frac{1}{2\sqrt{\sigma(x_1)}}\sum_{r \{\pm\}} \frac{1}{(x_1-x_r)M'(x_r)} \Big( \sum_{h=0}^g \sum_{J \subset K}^{'} \mathcal{W}_h(x_r,x_J) \mathcal{W}_{g-h}(x_r,x_{K/J}) \cr
&&+ W_{g-1}(x_r,x_r,x_K) +\gamma \left(\partial_x W_{g-1/2}(x,x_K) \right)_{|x=x_r}\Big)\cr
&&-\frac{1}{2y(x_1)} \Big( \sum_{h=0}^g \sum_{J \subset K}^{'} \mathcal{W}_h(x_1,x_J) \mathcal{W}_{g-h}(x_1,x_{K/J}) \cr
&&+ W_{g-1}(x_1,x_1,x_K) +\gamma \left(\partial_x W_{g-1/2}(x,x_K) \right)_{|x=x_1}\Big)\cr
\eea

In theory, one can implement the previous recursion to compute every order of every correlation functions. But in practice, the values of $W_g(x_\pm)$ are problematic since they involve taking the correct branch of the square root each time (and the choice is far from being trivial when the function gets complicated). Also implementing this recursion on a computer is very difficult because the square roots cannot be simplified easily in symbolic computations. For all these reasons, it seems more natural to get rid of the square roots by introducing a proper parametrization of the spectral curve. In particular in the one-cut case, the parametrization can be defined globally and will transform square root singularities into poles which are much easier to handle. 

\section{Solving the refined topological recursion with the Zhukovsky parametrization}

\subsection{Definition of a kernel function}

In the following sections, we will need the following function:
\beq 
S(z_1,z)=\frac{ \left(z-\frac{1}{z}\right)^2}{\left(z_1-\frac{1}{z_1}\right)\left(z-z_1\right)\left(z-\frac{1}{z_1}\right)}
\eeq
From the definition, it is straightforward to see that the function $z\to S(z_1,z)$ has the following properties:
\begin{enumerate} \label{prop kernel}
\item It satisfies the identity:
\beq \frac{1}{z^2} S(z_1,\frac{1}{z})=S(z_1,z)\eeq
\item The behavior at infinity is given by $S(z_1,z)\overset{z\to \infty}{\sim} \frac{1}{(z_1-\frac{1}{z_1})}+O\left(\frac{1}{z}\right)$
\item $z\to S(z_1,z)$ has two double zeros at $z=\pm 1$
\item $z\to S(z_1,z)$ has simple poles at $z=z_1$ and $z=\frac{1}{z_1}$ with residues respectively $1$ and $-1$. It also has a double pole at $z=0$ but no other singularities in the complex plane.
\end{enumerate}

As we will see in the next sections, this function $S(z_1,z)$ will be useful to solve the recursion. This function can be thought as an explicit formula for the ``recursion" kernel (often noted $dS_{z_1}(z)$) introduced in the solution of the topological recursion by Eynard and Orantin in \cite{OE}.

\subsection{Rewriting the loop equation with the Zhukovsky variable}

The main advantage of dealing with a one cut model is that there exists a convenient global parametrization of the spectral curve with a complex variable. This global parametrization with a complex variable is specific to the one-cut case, when there are several cuts, one has to deal with different patches of local parametrizations. In the case of hermitian matrix models, the reformulation of the solution by Eynard and Orantin in \cite{OE} only uses local properties around the branchpoints (extremities of the interval). This renders the explicit computation simple with standard algebro-geometric tools even for the multi-cut case. For arbitrary $\beta$, the situation is more complex since there is no known possibility to reformulate the recursion with only local expressions around the branchpoints. However, in the one-cut case, the global parametrization is sufficient to enable computations. We parametrize the spectral curve with the Zhukovsky variable:
\beq \label{referee2} x(z)=\frac{a+b}{2} +\frac{b-a}{4}\left(z+\frac{1}{z}\right)\eeq
It splits the complex plane into two different domains. Indeed since $x(\frac{1}{z})=x(z)$, it splits the complex plane into two regions: The first one is the outside of the unit disk where $|z|>1$ which we take as the ``physical sheet". The second region is the inside of the unit disk where $|z|<1$. The advantage of the $z$ variable is that the spectral curve can be rewritten in an non-ambiguous way:
\beq \tilde{y}(z)=y(x(z))=M(x(z)) \frac{b-a}{4} \left(z-\frac{1}{z}\right)=\tilde{M}(z) \left(z-\frac{1}{z}\right)\eeq
Here we introduce a tilde notation to distinguish the functions as being functions of $z$ compared to non-tilde functions being functions of $x$. This will avoid confusion when dealing with derivatives.

Then, we can express all correlation functions $W_g(x_1,\dots,x_n)$ as functions of the $z$ variable.
\beq \tilde{W}_g(z_1,\dots,z_n)=W_g(x(z_1),x(z_2),\dots,x(z_n))\eeq
In particular, the functions $\tilde{W}_g(z_1,\dots,z_n)$ are rational functions of the $z_i$ variables (the square roots disappear) which makes the computations rigorous and easier.
The next step is to rewrite the loop equations with the help of the $z$ function. We recall here the initial form of the loop equations \eqref{Projected loop}:
\bea  2y(x_1)W_g(x_1,x_K)&=& -\sum_{h=0}^g\sum_{I \subset K}^{'} \mathcal{W}_h(x_1,x_I)\mathcal{W}_{g-h}(x_1,x_{K/I}) -W_{g-1}(x_1,x_1,x_K)\cr
&&-\gamma \partial_{x_1} W_{g-\frac{1}{2}}(x_1,x_K)+P_g(x_1,x_K)
\eea
where the sum $\underset{h=0}{\overset{g}{\sum}}\underset{I \subset K}{\overset{'}{\sum}}$ means that we exclude the two case $(h=0, I=\emptyset)$ and $(h=g,I=K)$ which are already included in the l.h.s. Also remind that $x_1\mapsto P_g(x_1,x_K)$ is regular on the cut. We want to insert the $z$ variable, the only difficulty concerns the derivative relatively to $x$, which is not present in the $\beta=1$ case. Using the chain rule, we find:
\bea \label{Loop z} 2\tilde{y}(z_1)\tilde{W}_g(z_1,z_K)&=& -\sum_{h=0}^g\sum_{I \subset K}^{'} \tilde{\mathcal{W}}_h(z_1,z_I)\tilde{\mathcal{W}}_{g-h}(z_1,z_{K/I}) -\tilde{W}_{g-1}(z_1,z_1,z_K)\cr
&&-\frac{4z_1^2}{(b-a)(z_1^2-1)}\gamma \partial_{z_1} \tilde{W}_{g-\frac{1}{2}}(z_1,z_K)+\tilde{P}_g(z_1,z_K)
\eea
Note that since the function $x_1\to P_g(x_1,x_K)$ is regular at the cut (and thus has no square roots singularity), it implies that \beq\label{Regularity} \encadremath{\tilde{P}_g(z_1,z_K)=\tilde{P}_g(\frac{1}{z_1},z_K)}\eeq
an important property that will be used in the next section.

\subsection{Definition of several integration contours}

In the next section, we will need to use the following contours. $\mathcal{C}_+$ denotes a contour circling around the unit circle by its external face, whereas $\mathcal{C}_-$ is a contour lying inside the unit disk and circling the unit circle. Observe also that by contour deformation we have that a function $f(z)$ having only poles at $z=\pm 1$ along the unit circle will satisfy:
\beq \label{tarace} \Res_{z \to \pm 1} f(z)dz=\oint_{\mathcal{C}_-}f(z)dz- \oint_{\mathcal{C}_+}f(z)dz\eeq
With the help of these contours and some little complex analysis, we will see in the next section that the recursion \eqref{Loop z} can be solved explicitly, even if the functions $\tilde{P}_g(z_1,z_K)$ are unknown.

\subsection{Solving the recursion in the $z$ variable}

We now use the different properties of the kernel $S(z_1,z)$ in order to rewrite a solution of the loop equations as a residue formula. We also remind the reader that, by definition, the correlation functions satisfy the following properties:
\begin{enumerate}
\item $\td{W}_g(z, z_K)$ behaves like $O\left(\frac{1}{z^2}\right)$ as $z$ goes to infinity.
\item $\td{W}_g(z, z_K)$ has poles at $z=\pm 1$ and is a symmetric function
\item $\td{W}_g(z, z_K)$ is regular outside the unit disk
\end{enumerate}
These properties come directly from the definition of the correlation functions and the choice of the physical sheet to be the outside of the unit disk. However, as mentioned in \cite{Nadal}, the main difference with the hermitian case is that we no longer have the mirror relation:
\beq \td{W}_g(\frac{1}{z},z_K)=z^2\td{W}_g(z,z_K)\eeq
because of the derivative term (this is the reason why we have to take a kernel satisfying a similar property) in the recursion when $\beta\neq 1$. This mirror relation was fundamental to get the properties that the functions $\td{W}_g(z,z_K)$ are regular inside the unit disk (they are by definition regular outside the unit disk and it is extended inside the units disk by the map $z \to \frac{1}{z}$) and so that their only poles were at the branchpoints. For arbitrary $\beta$ since this mirror relation is no longer satisfied, the correlation functions may have poles at the zeros of $\td{M}(z)$ located inside the unit disk.

However one can perform the following transformations: First, we use properties of the kernel \eqref{prop kernel} and the fact that $\td{W}_g(z,z_K)$ is regular outside the unit disk and decreases as $O\left(\frac{1}{z^2}\right)$ at $z=\infty$. For every $z_1$ lying in the physical sheet (i.e. $|z_1|>1$) we have:
\beq \td{W}_g(z_1,z_K)=\Res_{z\to z_1} S(z_1,z)\td{W}_g(z,z_K)dz= -\frac{1}{2i\pi}\oint_{\mathcal{C}_+} S(z_1,z)\td{W}_g(z,z_K)dz\eeq
Then we use the fact that it satisfies the loop equations \eqref{Loop z}. In order to have more compact notations we define:
\bea \label{Rec} \text{Rec}_g(z,z_K)&=&-\sum_{h=0}^g\sum_{I \subset K}^{'} \tilde{\mathcal{W}}_h(z,z_I)\tilde{\mathcal{W}}_{g-h}(z,z_{K/I}) -\tilde{W}_{g-1}(z,z,z_K)\cr
&&-\frac{\gamma}{x'(z)} \partial_{z} \tilde{W}_{g-\frac{1}{2}}(z,z_K)
\eea
so that the loop equations are rewritten like:
\beq \label{Loop z 2} 2y(z)\td{W}_g(z,z_K)=\text{Rec}_g(z,z_K)+\td{P}_g(z,z_K)\eeq
Hence:
\beq \label{one}\td{W}_g(z_1,z_K)=-\frac{1}{2i\pi}\oint_{\mathcal{C}_+} \frac{S(z_1,z)}{2\td{y}(z)}\left(\text{Rec}_g(z,z_K)+\td{P}(z,z_K)\right)dz\eeq
We then perform a change of variable $z\to \frac{1}{z}$ inside the integral. The contour of integration is changed to $-\oint_{\mathcal{C}_-}$. Moreover the spectral curve satisfies $\td{y}(\frac{1}{z})=-\td{y}(z)$ whereas the function $\td{P}(z,z_K)$ satisfies $\td{P}(\frac{1}{z},z_K)=\td{P}(z,z_K)$ \eqref{Regularity}. Finally using the first property of \eqref{prop kernel} we get to:
\bea \label{two} \td{W}_g(z_1,z_K)&=&-\frac{1}{2i\pi}\oint_{\mathcal{C}_+} \frac{S(z_1,z)}{2\td{y}(z)}\left(\text{Rec}_g(z,z_K)+\td{P}(z,z_K)\right)dz\cr
&=&\frac{1}{2i\pi}\oint_{\mathcal{C}_-} \frac{S(z_1,z)}{2\td{y}(z)}\left(\text{Rec}_g(\frac{1}{z},z_K)+\td{P}(z,z_K)\right)dz\cr
\eea
Now, we sum \eqref{one} and \eqref{two} to get:
\bea 2\td{W}_g(z_1,z_K)&=&-\frac{1}{2i\pi}\oint_{\mathcal{C}_+} \frac{S(z_1,z)}{2\td{y}(z)}\left(\text{Rec}_g(z,z_K)+\td{P}(z,z_K)\right)dz\cr
&&+\frac{1}{2i\pi}\oint_{\mathcal{C}_-} \frac{S(z_1,z)}{2\td{y}(z)}\left(\text{Rec}_g(\frac{1}{z},z_K)+\td{P}(z,z_K)\right)dz\cr
&=& -\frac{1}{2i\pi}\oint_{\mathcal{C}_+} \frac{S(z_1,z)}{2\td{y}(z)}\text{Rec}_g(z,z_K)dz+\frac{1}{2i\pi}\oint_{\mathcal{C}_-} \frac{S(z_1,z)}{2\td{y}(z)}\text{Rec}_g(\frac{1}{z},z_K)dz\cr
&&+\frac{1}{2i\pi}\oint_{\mathcal{C}_- - \mathcal{C}_+} \frac{S(z_1,z)}{2\td{y}(z)}\td{P}_g(z,z_K)dz\cr
\eea

We now observe that the last contour circles the unit circle as explained in \eqref{tarace}. But the function $\td{P}_g(z,z_K)$ is regular along the cut and therefore does not have pole at $z=\pm 1$. The denominator $\td{y}(z)$ brings a simple pole at $z=\pm 1$, but since $S(z_1,z)$ has double zeros there, the function $z\to \frac{S(z_1,z)}{2\td{y}(z)}\td{P}_g(z,z_K)$ has no singularity along the unit circle giving a null integral. Hence:
\beq 2\td{W}_g(z_1,z_K)=-\frac{1}{2i\pi}\oint_{\mathcal{C}_+} \frac{S(z_1,z)}{2\td{y}(z)}\text{Rec}_g(z,z_K)dz+\frac{1}{2i\pi}\oint_{\mathcal{C}_-} \frac{S(z_1,z)}{2\td{y}(z)}\text{Rec}_g(\frac{1}{z},z_K)dz \eeq

\textbf{In particular, by the use of this contour, we have got rid of the unknown function $\td{P}_g(z,z_K)$ giving us the opportunity to compute recursively the correlation functions.}

In order to simplify the last expression, we perform the change of variable $z\to \frac{1}{z}$ only in the integral involving $\text{Rec}_g(\frac{1}{z})$. Again using the properties of the kernel $S(z_1,z)$, the fact that $\oint_{\mathcal{C}_-} \to -\oint_{\mathcal{C}_+}$ and that $\td{y}(\frac{1}{z})=-\td{y}(z)$ we find that:

\beq 
\frac{1}{2i\pi}\oint_{\mathcal{C}_-} \frac{S(z_1,z)}{2\td{y}(z)}\text{Rec}_g(\frac{1}{z},z_K)dz=-\frac{1}{2i\pi}\oint_{\mathcal{C}_+} \frac{S(z_1,z)}{2\td{y}(z)}\text{Rec}_g(z,z_K)dz
\eeq
so that in the end:
\beq \encadremath{ \label{Formule generale}
\td{W}_g(z_1,z_K)=-\frac{1}{2i\pi}\oint_{\mathcal{C}_+} \frac{S(z_1,z)}{2\td{y}(z)}\text{Rec}_g(z,z_K)dz }
\eeq

The last formula is indeed in total agreement with \cite{Marino} (noted with an index $_\text{BMS}$) where we have the equivalences:
\beq \td{y}(z)=-2y_{\text{BMS}}(x(z)) \,\,\,\, ,\,\,\,\, S(z_1,z)=-dS_{\text{BMS}}(x(z_1),x(z))x'(z)\eeq

Finally, the last step is to rewrite the last integrals as residue formulas. Observe that from the definition of $\text{Rec}_g(z,z_K)$ \eqref{Rec} it has no pole outside the unit circle except for the terms involving $\td{\mathcal{W}}_0(z,z_j)$ which has a double pole at $z=z_j$ (since the shifted version of the two points function is not regular). Indeed, it it just products/sums/derivatives of correlation functions of lower orders, which have no pole outside the unit circle. Moreover, $\text{Rec}_g(z,z_K)$ behaves well at infinity: $\text{Rec}_g(z,z_K)=O\left(\frac{1}{z^2}\right)$ (or even with higher powers). Thus, if we look at the part involving the integral over $\mathcal{C}_+$, we see that outside the unit disk, the function $z\to \frac{S(z_1,z)}{2\td{y}(z)}\text{Rec}_g(z,z_K)$ only has poles at $z=z_1$ (which is in the physical sheet, i.e. lies outside the unit disk) and $z=z_\pm$, the other zeros of the spectral curve lying outside the unit disk. Thus, we get (using \eqref{twopoints}):
\bea
\td{W}_g(z_1,z_K)&=&-\frac{1}{2i\pi}\oint_{\mathcal{C}_+} \frac{S(z_1,z)}{2\td{y}(z)}\text{Rec}_g(z,z_K)dz\cr
&=&\Res_{z\to z_1,z_{\pm}, z_K} \frac{S(z_1,z)}{2\td{y}(z)}\text{Rec}_g(z,z_K)dz\cr
&=& \frac{\text{Rec}_g(z_1,z_K)}{2\td{y}(z_1)} +\sum_{i \in \{\pm\}} \frac{ S(z_1,z_i)\text{Rec}_g(z_i)}{2\td{y}'(z_i)}\cr
&&+ \sum_{j=2}^K \partial_z \left(\frac{S(z_1,z)\td{W}_g(z,z_{K/\{j\}})}{2\td{y}(z)}\frac{8z^2z_j^2}{(b-a)^2\left(1-zz_j\right)^2}\right)_{|z=z_j}\cr
\eea
So that in the end:

\begin{equation*} \label{Final formula}
\addtolength{\fboxsep}{10pt}
			\boxed{
				\begin{split}
				\td{W}_g(z_1,z_K)=&\frac{\text{Rec}_g(z_1,z_K)}{2\td{y}(z_1)} +\sum_{i \in \{\pm\}} \frac{ S(z_1,z_i)\text{Rec}_g(z_i)}{2\td{y}'(z_i)}\cr
				&+ \sum_{j=2}^K \partial_z \left(\frac{S(z_1,z)\td{W}_g(z,z_{K/\{j\}})}{\td{y}(z)}\frac{8}{(b-a)^2\left(1-\frac{1}{zz_j}\right)^2}\right)_{|z=z_j}\cr
				\end{split}
			}
	\end{equation*}

This formula is very similar to the standard formula in matrix models as for example the one given in \cite{Marino} or \cite{Nadal}. Especially, formula \eqref{Formule generale} is still valid whatever spectral curve we have to deal with, as long as it only has one cut. For example, this can be used in Chern-Simons integrals, or its Stieljes-Wigert equivalent where the potential is of the form $\ln(x)^2$. Only the last part requires the exact knowledge of the form of the spectral curve, that is to say in our case that $M(x)$ is a polynomial of degree $2$. It is also worth noticing that all the formulas presented here are valid for every value of $\beta$ and can be easily implemented on a computer, since they only require residue computations.

\section{Application to the first correlation functions}

\subsection{Computation of $\td{W}_{1/2}(z)$}

The first step to implement the results of the former section is to compute $\td{W}_{1/2}(z)$, giving us the first correction to the one-point correlation function. In this case, the loop equation is:
\beq 
\td{W}_{1/2}(z_1)=\frac{\gamma}{2i\pi}\oint_{\mathcal{C}_+} \frac{S(z_1,z)}{2\td{y}(z)}(\partial_x W_0(x))_{| x=x(z)}dz
\eeq 
The link between $W_0(x)$ and the spectral curve is given by:
\beq W_0(x)=y(x)+\frac{1}{2}V'(x)\eeq

In the end by deforming the contour we get to:
\beq 
\encadremath{ \td{W}_{1/2}(z_1)=-\gamma\frac{\left(y'(x)+\frac{1}{2}V''(x)\right)_{x=x(z_1)}}{2\td{y}(z_1)} -\gamma\sum_{i \in \{\pm\}} \frac{ S(z_1,z_i) \left(y'(x)+\frac{1}{2}V''(x)\right)_{x=x(z_i)}}{2\td{y}'(z_i)} }
\eeq 

Another possible expression for this correlation function is to get rid of the potential sooner in the computation. Indeed, since $V''(x)$ is a function of $x$, it satisfies $V''(x)_{|x=x(z)}=V''(x)_{|x=x(1/z)}$ and it is regular in the whole complex plane. Therefore it can be eliminated in the same way as the unknown function $\td{P}_{1/2}(z)$ in the former section. However, doing so has a main drawback, specifically that if the formula:
 \bea 
\td{W}_{1/2}(z_1)&=&\frac{\gamma}{2i\pi}\oint_{\mathcal{C}_+} \frac{S(z_1,z)}{2\td{y}(z)}(\partial_x y(x))_{| x=x(z)}dz\cr
&=&\frac{\gamma}{2i\pi}\oint_{\mathcal{C}_+} \frac{S(z_1,z)\td{y}'(z)}{2x'(z)\td{y}(z)}dz
\eea
is still valid, then when deforming the contour, we get a simple pole at $z=\infty$ and therefore a residue there. This residue can be computed easily from the leading behavior of all the factors at infinity. We find that it is given by (note that $\frac{\td{y}'(z)}{\td{y}(z)}\sim \frac{3}{z}$ in our case of a quartic potential and is more generally $\frac{d-1}{z}$ for a polynomial potential of degree $d$):
$$\Res_{z=\infty} \frac{S(z_1,z)\td{y}'(z)}{2x'(z)\td{y}(z)}dz= -\frac{3}{2\frac{(b-a)}{4}(z_1-\frac{1}{z_1})}$$
Therefore, the first correction can also be expressed as:
\beq \label{Formule W_1/2}
\encadremath{ \td{W}_{1/2}(z_1)=\frac{6\gamma}{(b-a)(z_1-\frac{1}{z_1})}-\frac{\gamma \td{y}'(z_1)}{2x'(z_1)\td{y}(z_1)}- \gamma\sum_{i \in \{\pm\}}\frac{S(z_1,z_i)}{2x'(z_i)} }
\eeq 

This last formula is similar to eq. (2.53) given in \cite{Marino} with $d=4$.

\subsection{Leading order of the two-points correlation function}

In order to start the recursion, one also need to get the function $\td{W}_0(z_1,z_2)$ which is not given by the standard recursion \eqref{Loop z}. However, from the work in the $x$ variable \eqref{ouh} it is very straightforward to get it:

\beq \encadremath{ \td{W}_0(z_1,z_2)= \frac{1}{x'(z_1)x'(z_2)}\frac{1}{(z_1z_2-1)^2}=\frac{16z_1^2z_2^2}{(b-a)^2(z_1^2-1)(z_2^2-1)(z_1z_2-1)^2} }\eeq
Note that it is regular at $z_1=z_2$ which is not the case for its shifted version (the one needed in the recursion):
 \beq \encadremath{ \tilde{\mathcal{W}}_0(z_1,z_2)=\frac{16z_1^2z_2^2}{(b-a)^2(z_1z_2-1)^2} \left(\frac{1}{(z_1^2-1)(z_2^2-1)}+\frac{1}{2(z_1-z_2)}\right) }\eeq

Finally, note that both $\tilde{W}_0(z_1,z_2)$ and $\tilde{\mathcal{W}}_0(z_1,z_2)$ have double poles with residues at $z_1=\frac{1}{z_2}$. The behavior of the shifted version $\tilde{\mathcal{W}}_0(z_1,z_2)$ around $z_1=z_2$ is given by:
\beq \label{twopoints}\tilde{\mathcal{W}}_0(z_1,z_2)=\frac{8}{(b-a)^2(z_1-z_2)^2\left(1-\frac{1}{z_1z_2}\right)^2}+ \text{Reg}_{z_1\to z_2}\eeq

\subsection{Computing $\td{W}_1(z_1)$}

The next important correlation function to compute is $\td{W}_1(z_1)$ which will give the second correction to the one-point correlation function. For arbitrary $\beta$, the loop equations give us that:

\beq \text{Rec}_1(z)=-\td{W}_0(z,z)-\frac{\gamma}{x'(z)}\partial_z \td{W}_{1/2}(z)\eeq

Note that in this formula we have to use the non-shifted version of the leading order of the two-points function, which is indeed regular when its two variables coincide. In particular, we have:
\beq \td{W}_0(z,z)=\frac{1}{x'(z)^2(z^2-1)^2}=\frac{16 z^4}{(b-a)^2(z^2-1)^4}\eeq
To get the term involving $\td{W}_{1/2}(z)$, one can use \eqref{Formule W_1/2}. Finally we find that:

\bea \label{Formule W_1}\td{W}_1(z_1)&=& -\frac{1}{2\td{y}(z_1)x'(z_1)^2(z_1^2-1)^2}-\sum_{i \in \{\pm\}} \frac{ S(z_1,z_i)}{2x'(z_i)^2(z_i^2-1)^2\td{y}'(z_i)}\cr
 &&- \gamma\frac{\partial_{z_1} \td{W}_{1/2}(z_1)}{2x'(z_1)\td{y}(z_1)}-\gamma\sum_{i \in \{\pm\}} \frac{ S(z_1,z_i)\left(\partial_z \td{W}_{1/2}(z)\right)_{|z=z_i}}{2x'(z_i)\td{y}'(z_i)}\cr
\eea

Note that there are two different contributions: only the first one remains when $\beta=1$ (i.e. $\gamma=0$) whereas the second line of the last equation introduces corrections proportional to $\gamma^2$ (Remember that $\td{W}_{1/2}(z)$ is already proportional to $\gamma$).

\subsection{Computation of $\td{W}_{1/2}(z_1,z_2)$}

In order to verify our algorithm for computing the correlation functions, it is interesting to compute $\td{W}_{1/2}(z_1,z_2)$ and to verify that it is indeed a symmetric function of its variables (a fact that is far from trivial from the definition of the algorithm). In this case, the recursion scheme is:
\beq \text{Rec}_{1/2}(z,z_2)=-\frac{\gamma}{x'(z)} \partial_z \td{W}_0(z,z_2)-2\td{\mathcal{W}}_0(z,z_2)\td{W}_{1/2}(z)\eeq
so that $\td{W}_{1/2}(z_1,z_2)$ is given by:
\bea \td{W}_{1/2}(z_1,z_2)&=&-\frac{\gamma}{2x'(z_1)\td{y}(z_1)} \partial_{z_1} \td{W}_0(z_1,z_2)-\frac{\td{\mathcal{W}}_0(z_1,z_2)\td{W}_{1/2}(z_1)}{\td{y}(z_1)}\cr
&&-\gamma \sum_{i \in \{\pm\}}\frac{ S(z_1,z_i)\left(\partial_z \td{W}_0(z,z_2)\right)_{|z=z_i}}{2x'(z_i)\td{y}'(z_i)}
-\sum_{i \in \{\pm\}}\frac{ S(z_1,z_i)\td{\mathcal{W}}_0(z_i,z_2)\td{W}_{1/2}(z_i)}{\td{y}'(z_i)}\cr
&&-\partial_z \left(\frac{S(z_1,z)\td{W}_{1/2}(z)}{\td{y}(z)}\frac{8}{(b-a)^2\left(1-\frac{1}{zz_2}\right)^2}\right)_{|z=z_2}\cr
\eea

As mentioned before, it is rather non-trivial from such a definition that $\td{W}_{1/2}(z_2,z_1)=\td{W}_{1/2}(z_1,z_2)$, but it can easily be verified on examples on a computer. Note also that all the terms in the last formula are all exactly proportional to $\gamma$ as expected.

\section{Getting information back to the density of eigenvalues}

One of the aims to this article is to provide simulations for the density of eigenvalues and especially to get the first corrections for arbitrary value of $\beta$. In this context, it is necessary to get the information back from the correlation functions $\td{W}_{1/2}(z)$ and $\td{W}_1(z)$ to the density of eigenvalues. Using the change of variable \eqref{referee2} and $z = e^{i\theta}$, the density of eigenvalues expressed in the variable $\theta$ (keeping into account the Jacobian $\frac{dx}{d\theta}=\frac{b-a}{2}\sin\theta$) can be written as follows:
\beq \rho_\infty(\theta)= \frac{1}{2i\pi t_0}\frac{(b-a)}{2}\sin\theta \left(\td{W}_0(e^{-i\theta})- \td{W}_0(e^{i\theta})\right)\eeq 
Inserting the next $\frac{1}{N}$ corrections has to be done with caution. Indeed, in general $W_g(z)$ may have poles of high order at $z=\pm1$ so that it is unsure that the combination $\rho_g(\theta)=\frac{1}{2i\pi}\frac{(b-a)}{2N^g}\sin\theta\left( \td{W}_{g}(e^{-i\theta})-\td{W}_{g}(e^{i\theta})\right)$ will not be singular at $\theta\in \{0,\pi\}$ (but taking this combination may also decrease the order of the singularity). From the theory we expect a weak convergence that is to say that for any test function $\phi(\theta)$, the integrals $\frac{1}{N^g}\int_0^\pi \rho_g(\theta) \phi(\theta)d\theta$ should give better and better refinements to the observed $\int_0^\pi \rho^{(N)}(\theta)\phi(\theta)d\theta$ for a given and finite $N$. If $\rho_g(\theta)$ is integrable, then the space of test functions can be chosen as continuous functions, but if it is not integrable, then the space of test functions have to be restricted to compactly supported functions within $(0,\pi)$. We let this discussion open here since in the example we study, the first correction $\rho_\frac{1}{2}(\theta)$ is continuous functions on $[0,\pi]$ and the convergence can be studied graphically without having to compute the moments (i.e. taking $\phi(\theta)=x(e^{i\theta})^k$).

\section{Simulations}

We take a specific value of the potential $V(x)=x^4+\frac{1}{2}x^2$. This potential as a unique minimum on the real axis at $x=0$ and we therefore expect a one-cut solution around $x=0$. Moreover, since the potential is even, all correlation functions are expected to be even, that is to say that the final density of angles should exhibit a symmetry around $\frac{\pi}{2}$. In particular, the extremities of the interval always satisfy $b=-a$ and are given by:
\beq \frac{3}{4}a^4+\frac{1}{4}a^2=t_0 \Leftrightarrow b=\frac{\sqrt{-6+6\sqrt{1+48t_0^2}}}{6}=-a\eeq
which is independent of $\beta$ as expected. In this case, the limiting density of eigenvalues is given by:
\beq \rho_\infty(\theta)= \frac{1}{\pi t_0^2}\left(2b^2\cos^2\theta+b^2+\frac{1}{2}\right)b^2\ sin^2\theta\eeq
This last formula depends on the parameter $t_0$ in the following picture:

\begin{center} 
	\includegraphics[height=6cm]{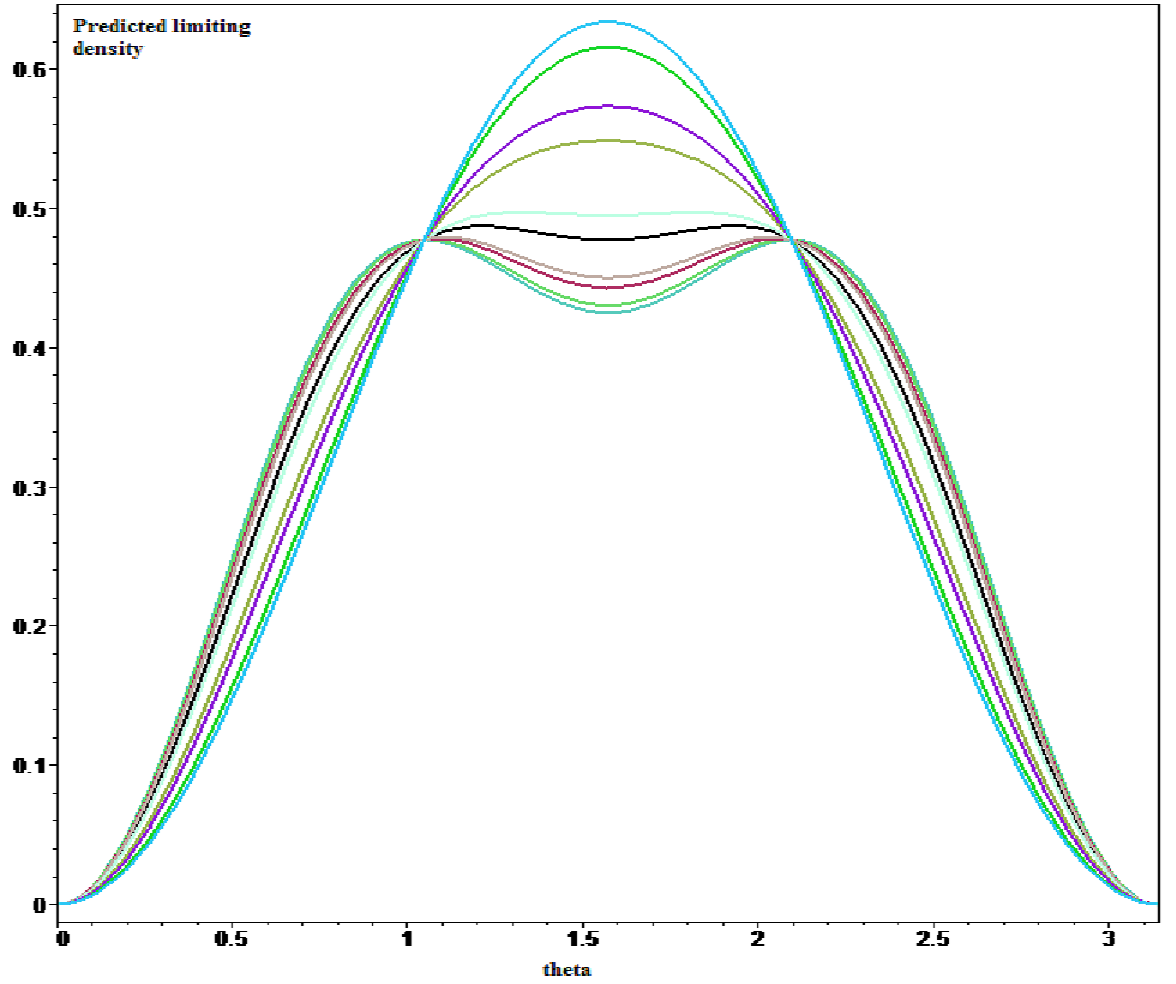}
	
	\textit{\underline{Fig.3}: Different limiting densities regarding the value of the parameter} $t_0$\textit{. The values taken are (from top to bottom)} $10^{-3}, 10^{-2}, 0.05, 0.1, 0.5, 1$ \textit{(in black) and} $5, 10, 100, 10^4$
\end{center} 

Moreover, a straightforward computation gives us:
\beq \label{rrrrrrrrr}\rho_\frac{1}{2}(\theta)= -\frac{\gamma}{4\pi}\left(4-\frac{2(z_+^2-1)}{z_+^2-2z_+\cos(\theta)+1} -\frac{2(z_-^2-1)}{z_-^2-2z_-\cos(\theta)+1}\right)\eeq
where we remind the reader that $z_+$ and $z_-$ are the two solutions of $\td{M}(z)=0$ located outside the unit circle. In particular we note that even if $W_\frac{1}{2}(z)$ has some poles at $z=\pm 1$, taking the combination $W_\frac{1}{2}(\frac{1}{z})- W_\frac{1}{2}(z)$ only has a $\frac{1}{\sin\theta}$ singularity that cancels when multiplying by the Jacobian to get the density. It is expected (Cf. \cite{Referee3} for a discussion about subleading terms of the density) that subleading densities should, if integrable, have a zero total integral and hence cannot be positive. In our situation, the result is immediate from formula \eqref{rrrrrrrrr} since we have by symmetry $z_-=-z_+$ with $z_+$ outside the unit circle. The antiderivative can be computed explicitly and the integral for $\theta\in[0,\pi]$ is automatically zero when $z_+$ is outside the unit circle.
 
One can compare these limiting distributions with different histograms of eigenvalues obtained for simulations (we use a standard Metropolis algorithm which is detailed in \cite{These}) with different values of $\beta$.

\begin{center}
  \includegraphics[height=9cm]{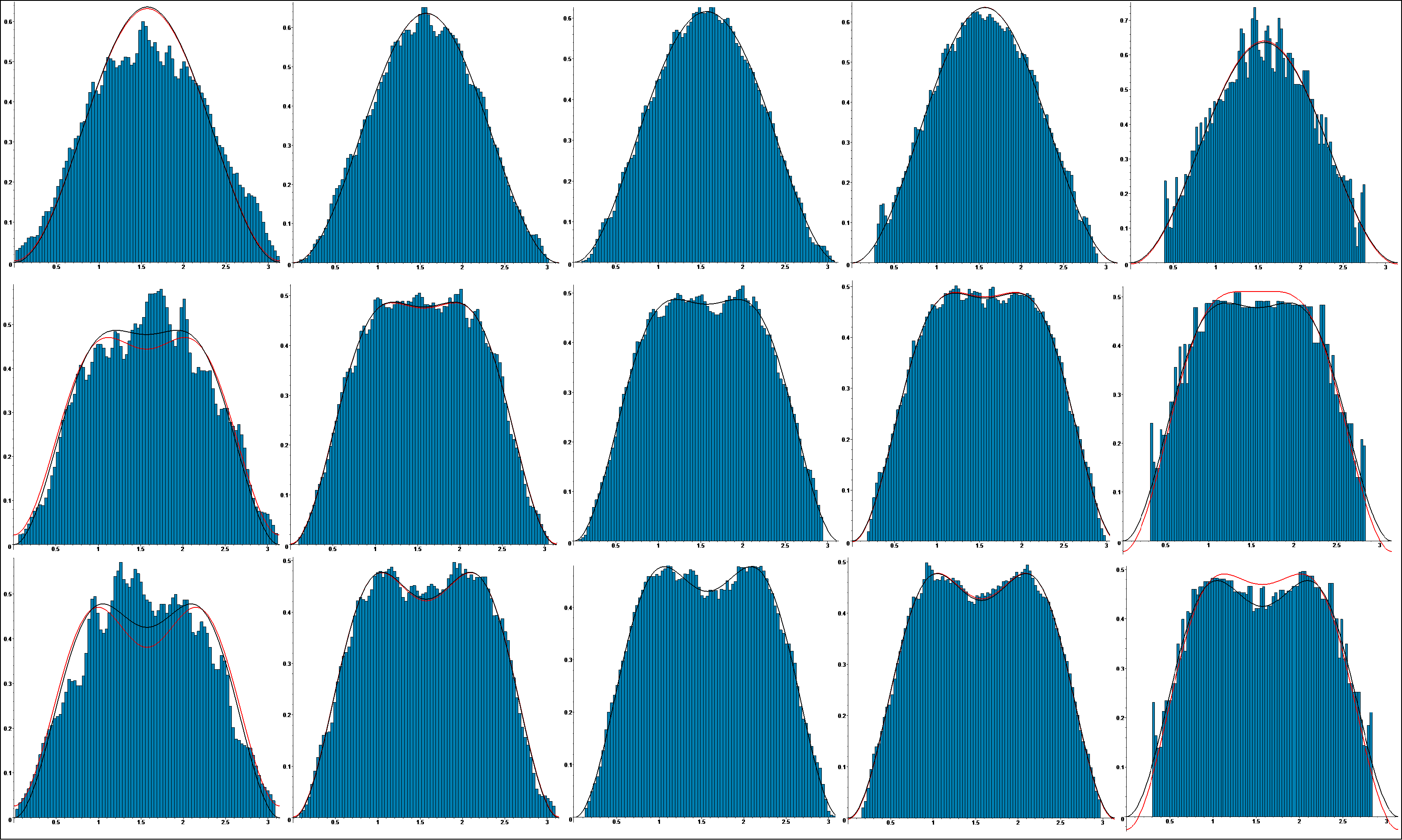}

\textit{\underline{Fig. 4}: Different simulations with $N=50$. The black line represents the theoretical limiting eigenvalues density (independent of $\beta$). The value of $t_0$ is respectively $10^{-2}$, $1$ and $100$ in the first, second and third rows. In each rows the value of the $\beta$ parameter is respectively from left to right: $10^{-2}$, $\frac{1}{2}$, $1$, $2$, $100$. In red is the first correction in $\frac{1}{N}$ of the density computed with \eqref{rrrrrrrrr}} 
\end{center}

The conclusions of the simulations are the following. First, for $N=50$, we can see a very good agreement between the theoretical limiting density and the simulations when $\beta$ is close to the hermitian case $1$ and this independently of the value of $t_0$ which affects the curve. Secondly, we observe that when $\beta$ gets far from $1$, the theoretical limiting density is no longer a perfect approximation. However if we include the first correction in $\frac{1}{N}$ it gives a better (but not perfect) approximation of the observed density, especially regarding the endpoints. 

\section{Acknowledgments}

The author is especially grateful to V. Bouchard for useful discussions and giving an opportunity to continue theoretical research. We also want to thank B. Eynard and G. Borot for email correspondences. This work is supported by MS081 start-up grant offered by the University of Alberta and the government of Canada. The author is also grateful to the referee for suggesting improvements in the paper. Finally, the author address special thanks to Audrey, Didier, Colette, Julien, David, C\'{e}cilie and Sophie for giving him support and visiting him during the cold and never-ending Albertan winter.


\begin{thebibliography}{99}
\bibliographystyle{plain}

\bibitem{Nadal} G. Borot, B. Eynard, S. N. Majumdar, C. Nadal, ``Large deviations of the maximal eigenvalue of random matrices", \textit{arXiv}:10009.1945v3 [math-ph], 2010.

\bibitem{Marino} A. Brini, M. Mari\~{n}o, S. Stevan,  ``The uses of the refined matrix model recursion", \textit{arXiv}:1010.1210v2 [math-ph], 2010.

\bibitem{OE} B. Eynard, N.Orantin ``Invariants of algebraic curves and topological expansion", \textit{Comm. in Number Theory and Physics}, Vol \textbf{1}, No 2, 2007.

\bibitem{Marino2} M. Mari\~{n}o, ``Les Houches lectures on matrix models and topological strings", \textit{arXiv}:hep-th/0410165v3, 2004.

\bibitem{ChekEynbeta} L. Chekhov, B. Eynard, ``Matrix eigenvalue model: Feynman graph technique for all genera", \textit{JHEP}, Vol \textbf{26}, 2006.

\bibitem{Correct} B. Eynard, ``Topological expansion for the 1-hermitian matrix model correlation functions", \textit{JHEP} \textbf{0411}, 031, 2004.

\bibitem{Tricomi} F.G. Tricomi, ``Integral Equations", \textit{Dover Publications Inc.} (New edition, London 1985), 238 pages, ISBN 978-0486648286.

\bibitem{Referee1}  N. Ercolani, K. McLaughlin
``Asymptotics of the partition function for random matrices via Riemann-Hilbert techniques and applications to graphical enumeration", \textit{IMRN}, no. \textbf{14}, 755-820, 2003.

\bibitem{Referee3} K. Johansson ``On fluctuation of eigenvalues of random Hermitian matrices", \textit{Duke Math. J.}, Vol \textbf{91}, no. 1, 151–204, 1998. 

\bibitem{MultiCutCase} G. Bonnet, F. David, B. Eynard, ``Breakdown of universality in the multi-cut matrix models", \textit{J. Phys. A: Math.} Gen. 33 \textbf{6739}, 2000. 

\bibitem{MultiCutCase2} L. Lenoble, L. Pastur, ``On the asymptotic behaviour of correlators of multi-cut matrix models", \textit{J. Phys. A: Math.} Gen. 34 \textbf{L409}, 2001. 

\bibitem{ChekEynFg} L. Chekhov, B. Eynard, ``Hermitian matrix model free energy: Feynman graph technique for all genera", \textit{JHEP}, Vol \textbf{03}, 2006.

\bibitem{Eguchi} T. Eguchi, K. Maruyoshi, ``Penner Type Matrix Model and Seiberg-Witten Theory", \textit{JHEP} 1002 \textbf{022}, 2010.

\bibitem{Hardedges} B. Eynard, ``Loop equations for the semiclassical 2-matrix model with hard edges",  \textit{J. Stat. Mech.}, \textbf{P10006}, 2005.

\bibitem{Hardedges2} L. Chekhov, ``Solving matrix models in the 1/N-expansion", \textit{Russ. Math. Surv.}, Vol \textbf{61}, No 3, 2006.

\bibitem{Mironov} A. Mironov, S. Shakirov, ``The matrix model version of AGT conjecture and CIV-DV prepotential", \textit{JHEP} \textbf{1008} 066, 2010.

\bibitem{MoiBertrand} B. Eynard, O. Marchal, ``Topological expansion of the Bethe ansatz, and non-commutative algebraic geometry", \textit{JHEP} \textbf{0903} 094, 2009.

\bibitem{MoiLeonidBertrand} L. Chekov, B. Eynard, O. Marchal, ``Topological expansion of the Bethe ansatz, and quantum algebraic geometry" \textit{arXiv}:0911.1664v2 [math-ph], 2010.

\bibitem{MoiLeonidBertrand2} L. Chekov, B. Eynard, O. Marchal, ``Topological expansion of beta-ensemble model and quantum algebraic geometry in the sectorwise approach" \textit{arXiv}:1009.6007v1 [math-ph], 2010.

\bibitem{Dum} I. Dumitriu, A. Edelman, ``Matrix models for beta ensembles", \textit{arXiv}:0020.6043v1 [math-ph], 2002.

\bibitem{These} O. Marchal, ``Aspects g\'{e}om\'{e}triques et int\'{e}grables des mod\`{e}les de matrices al\'{e}atoires", PhD Thesis,	\textit{arXiv}:1012.4513v1 [math-ph], 2010.

\bibitem{BleherNotes} P. Bleher ``Lectures on random matrix models. The Riemann-Hilbert approach", \textit{arXiv}:0801.1858v2 [math-ph], 2008.

\bibitem{AGT} L. Alday, D. Gaiotto, Y. Tachikawa, ``Liouville Correlation Functions from Four-dimensional Gauge Theories", \textit{Letters in math. physics}, Vol \textbf{91}, No 2, 167-197, 2010.

\bibitem{Recent} G. Bonelli, K. Maruyoshi, A. Tanzini, ``Quantum Hitchin Systems via $\beta$-deformed Matrix Models", \textit{arXiv}:1104.4016v1 [hep-th], 2011.

\bibitem{Iqbal} A. Iqbal, C. Kozcaz, C. Vafa, ``The refined topological vertex", \textit{JHEP}, \textbf{0910}, 069, 2009.

\bibitem{Mehta} M. L. Mehta, ``Random matrices (3e edition)", \textit{Pure and Applied Mathematics Series} \textbf{142}, Elsevier (London - 2004), 688 pp. ISBN 0120884097.

\bibitem{Nekrasov} P. Sulkowski, ``Matrix models for $\beta$-ensembles from Nekrasov partition functions", \textit{JHEP} \textbf{1004}, 2010.

\bibitem{EMO} B. Eynard, M. Mari\~{n}o, N. Orantin, ``Holomorphic anomaly and matrix models", \textit{JHEP} \textbf{0706} 058, 2007.

\bibitem{Chekhov} L. Chekhov, ``Logarithmic potential beta-ensembles and Feynman graphs", \textit{arXiv}:1009.5940 [math-ph], 2009.

\end{thebibliography}
\end{document}